\documentclass[sigconf,nonacm]{acmart} 
\pdfoutput=1
\AtBeginDocument{%
  \providecommand\BibTeX{{%
    \normalfont B\kern-0.5em{\scshape i\kern-0.25em b}\kern-0.8em\TeX}}}
\makeatletter
\gdef\@copyrightpermission{
  \begin{minipage}{0.2\columnwidth}
   \href{https://creativecommons.org/licenses/by/4.0/}{\includegraphics[width=0.90\textwidth]{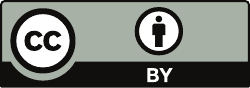}}
  \end{minipage}\hfill
  \begin{minipage}{0.8\columnwidth}
   \href{https://creativecommons.org/licenses/by/4.0/}{This work is licensed under a Creative Commons Attribution International 4.0 License.}
  \end{minipage}
}
\makeatother
\setcopyright{cc}
\setcctype[4.0]{by}
\acmConference[ScaleSys '25]{1st International Workshop on Intelligent and Scalable Systems Across the Computing Continuum}{November 19, 2025}{Vienna, Austria}
\acmBooktitle{1st International Workshop on Intelligent and Scalable Systems Across the Computing Continuum (ScaleSys '25), November 19, 2025, Vienna, Austria}
\usepackage[T1]{fontenc}
\usepackage{epstopdf}
\usepackage{graphicx}
\usepackage{placeins}
\usepackage{float}
\usepackage{nomencl}
\usepackage{balance} 
\usepackage{color,soul}
\usepackage{multirow}
\usepackage{makecell}
\usepackage{longtable}
\usepackage{hhline}
\usepackage{pifont}
\usepackage{tabularx}
\usepackage[inline,shortlabels]{enumitem}
\usepackage{geometry}
\usepackage{xspace}
\usepackage[binary-units,per-mode=symbol,exponent-product=\cdot,list-units=single,list-final-separator={,}]{siunitx}
\DeclareSIUnit[number-unit-product = ]\pixel{p}
\usepackage{chemformula}
\usepackage{gensymb}
\usepackage{amsmath}
\usepackage{booktabs}
\usepackage{comment}
\usepackage{footnote}
\usepackage{tablefootnote}
\usepackage{caption}
\usepackage{subcaption} 
\usepackage{wrapfig}
\usepackage[linesnumbered,ruled,vlined]{algorithm2e}
\usepackage[noend]{algpseudocode}
\usepackage{makecell}
\usepackage{tikz}
\usepackage[shortcuts]{extdash}
\usepackage{calc}
\DeclareRobustCommand{\numbercircle}[1]{%
  \tikz[baseline=(c.base)]{
    \node[draw,circle,minimum size=1.7em,inner sep=0pt] (c) {};
    \node at (c.center) {\sffamily\footnotesize #1};
  }%
}
\newlist{tasklist}{enumerate}{1}
\setlist[tasklist]{%
  labelsep=.6em,
  labelwidth=\widthof{\numbercircle{T4b}}, % usa l'etichetta più larga
  leftmargin=!,                              % forza l'uso di labelwidth
  align=parleft,
  itemsep=2pt, parsep=2pt, topsep=.4\baselineskip,
  listparindent=0pt
}
\usepackage{xurl}
\usepackage{orcidlink}
\usepackage{hyperref}
\usepackage{hyperxmp}
\usepackage{footmisc}
\hypersetup{pdfauthor=author}
\begin{document}
\title{Serverless Everywhere: A Comparative Analysis of\\ WebAssembly Workflows Across Browser, Edge, and Cloud}
% \author{Mario Colosi\textsuperscript{1}, Reza Farahani\textsuperscript{2}, Lauri Loven\textsuperscript{3}, Radu Prodan\textsuperscript{4}, Massimo Villari\textsuperscript{1}}

% \affiliation{
%   \vspace{0.7em}
%   \institution{\textsuperscript{1}MIFT Department, University of Messina, Italy \\%
%   \textsuperscript{2}Institute of Information Technology, University of Klagenfurt, Austria \\%
%   \textsuperscript{3}Center for Ubiquitous Computing, University of Oulu, Finland \\%
%   \textsuperscript{4}Department of Computer Science, University of Innsbruck, Austria}
% }
% --- ACM-required per-author entries (keeps visual look the same) ---
\author{Mario Colosi}
\affiliation{%
  \institution{MIFT Department, University of Messina}
  \city{Messina}
  \country{Italy}
}

\author{Reza Farahani}
\affiliation{%
  \institution{Institute of Information Technology, University of Klagenfurt}
  \city{Klagenfurt}
  \country{Austria}
}

\author{Lauri Lovén}
\affiliation{%
  \institution{Center for Ubiquitous Computing, University of Oulu}
  \city{Oulu}
  \country{Finland}
}

\author{Radu Prodan}
\affiliation{%
  \institution{Department of Computer Science, University of Innsbruck}
  \city{Innsbruck}
  \country{Austria}
}

\author{Massimo Villari}
\affiliation{%
  \institution{MIFT Department, University of Messina}
  \city{Messina}
  \country{Italy}
}

\renewcommand{\shortauthors}{Mario Colosi, Reza Farahani, Lauri Loven, Radu Prodan, Massimo Villari.}
%%%%%%%%%%%
\begin{abstract}
WebAssembly (Wasm) is a binary instruction format that enables portable, sandboxed, and near-native execution across heterogeneous platforms, making it well-suited for serverless workflow execution on browsers, edge nodes, and cloud servers. However, its performance and stability depend heavily on factors such as startup overhead, runtime execution model (e.g., Ahead-of-Time (AOT) and Just-in-Time (JIT) compilation), and resource variability across deployment contexts.
This paper evaluates a Wasm-based serverless workflow executed consistently from the browser to edge and cloud instances. The setup uses wasm32-wasi modules: in the browser, execution occurs within a web worker, while on Edge and Cloud, an HTTP shim streams frames to the Wasm runtime. We measure cold- and warm-start latency, per-step delays, workflow makespan, throughput, and CPU/memory utilization to capture the end-to-end behavior across environments. Results show that AOT compilation and instance warming substantially reduce startup latency. For workflows with small payloads, the browser achieves competitive performance owing to fully in-memory data exchanges. In contrast, as payloads grow, the workflow transitions into a compute- and memory-intensive phase where AOT execution on edge and cloud nodes distinctly surpasses browser performance.
%%%
\end{abstract}
\maketitle
\section{Introduction}
\label{sec:intro}
In the serverless paradigm, providers manage the scalability, placement, and lifecycle of short-lived, event-driven functions. While this abstraction significantly reduces the operational overhead for developers, it exposes drawbacks, including cold starts, queue latency, performance variability, and cost unpredictability~\cite{farahani2024serverless}. Optimizing these aspects requires consistent measurement methodologies across the computing continuum, i.e., browsers, edge instances, and cloud servers~\cite{farahani2024heftless}. Recent works have analyzed the causes of cold starts, offering a deeper understanding of their cost implications and underscoring the need for rigorous comparative evaluations~\cite{golec2024cold}.

WebAssembly (Wasm), with its compact bytecode, fast validation and instantiation, and strong sandboxed isolation, serves as a unifying execution substrate across the computing continuum~\cite{10.1145/3714465, farahani2023towards}. Recent advancements, such as the maturation of runtimes and the standardization of the \emph{WebAssembly System Interface} (WASI)~\footnote{\url{https://wasi.dev}}, have brought the ``\textit{compile once, run anywhere}'' paradigm closer to reality. In practice, the same Wasm artifact (i.e., a compiled and portable binary module) can be executed across diverse contexts without modification, offering two key benefits: \textit{(i)} methodological comparability and \textit{(ii)} a simplified software supply chain with uniform packaging and minimal system dependencies.

Recent studies confirm the competitiveness of Wasm compared to solutions like containers and Function-as-a-Service (FaaS) platforms, particularly under resource-constrained conditions~\cite{10787384}. However, trade-offs remain in terms of startup latency and memory footprint, indicating that Wasm’s advantages are not uniform across all deployment contexts. Comparisons between x86 and ARM architectures further demonstrate the growing maturity of this technology while revealing runtime variations with practical implications for system design and development choices~\cite{10701368}.
Nevertheless, existing evaluations are restricted to a single environment or micro-benchmark, offering limited insight into the comparison of end-to-end latency (i.e., function cold/warm behavior, function execution and communication time, and workflow makespan) for realistic application workflows. 

To our knowledge, no unified methodological work has yet executed the same Wasm-based workflow across browsers, edge nodes, and cloud platforms using consistent and homogeneous metrics. Thus, this paper presents a comparative analysis of serverless workflows implemented in Wasm and executed across three distinct environments: the browser, the edge, and the cloud. The entire evaluation is performed under a unified experimental setup, ensuring methodological consistency and enabling a fair, cross-environment comparison of performance and resource efficiency. The results of this analysis provide practical guidance for making informed orchestration decisions (e.g., function placement) in serverless workflows. The key contributions of this paper are:
\begin{enumerate}[(i)]
  \item we implement a serverless workflow in Wasm, exposing WASI interfaces compatible with three environments, browser, edge, and cloud;
  \item we isolate the effects of the execution environment and quantify workflow performance on cold and warm starts, function makespan and workflow makespan, throughput, and resource utilization (CPU, memory);
  \item we compare the performance of ahead-of-time (AOT) and just-in-time (JIT) as compilation strategies, along with a pre-warming method, to evaluate their impact within each environment.
\end{enumerate}

This paper has six sections: Section~\ref{sec:sota} summarizes the background and related work on Wasm, serverless computing, workflow processing, and cold start latency reduction. Section~\ref{sec:methodology} describes the strategy adopted for browsers, edge instances and cloud servers. Section~\ref{sec:setup} explains our evaluation setup before describing the experimental results in Section~\ref{sec:result}. Section~\ref{sec:conclusion} finally concludes the paper.
\section{Related Work}
\label{sec:sota}
This section reviews related work in three main areas: Wasm as an execution substrate for serverless platforms, serverless workflow orchestration strategies, and cold-start mitigation and cost modeling in serverless systems.
%%%%%%%%%%%%%%%%
\subsection{Wasm as a serverless execution substrate}
Recent measurements show that Wasm can serve as a competitive alternative to, or an integration layer with, containers~\cite{pham2023webassembly}. However, its maturity and compatibility still vary across runtime implementations and their integration with the underlying execution environment~\cite{10.1145/3712197}. Systematic comparison of Wasm runtimes ranks standalone and browser-integrated executors across execution models, interpreter, just-in-time (JIT), and ahead-of-time (AOT) compilation, as well as system interfaces and security aspects, highlighting significant differences in initialization latency and throughput~\cite{Talu2025}. These findings guide the selection of runtimes and configuration options, such as AOT and JIT compilation modes, as well as caching strategies evaluated in our experiments.

On the edge side, benchmarks show that Wasm-based serverless platforms can achieve reduced startup times and competitive efficiency profiles compared to widely used container-based solutions~\cite{mendki2020evaluating}. However, methodological differences in workloads, metrics, and network configurations across studies highlight the need for a homogeneous measurement protocol~\cite{10787384}. In parallel, recent analyses show 
the maturation of client-side, in-browser execution~\cite{de2021runtime,mohan2022comparative}. While Wasm offers advantages in portability and time-to-use for scientific computing, it remains constrained by limited tooling and restricted system access~\cite{perkel2024no}. Furthermore, in-browser large language model (LLM) inference systems demonstrate that compute-intensive operations can be executed directly on the client using WebGPU for acceleration and Wasm for CPU-bound computation, achieving near-native performance on the same hardware~\cite{10.1145/3696410.3714553,ruan2024webllm}.
\subsection{Serverless workflow orchestration strategies}
On the other hand, existing research on serverless workflow orchestration investigates the effects of placement strategies on workflow makespan and resource consumption~\cite{farahani2025energyless, ristov2023large, farahani2024heftless}. Edge resource allocation mechanisms using serverless platforms typically employ formalized policies derived from observable runtime metrics~\cite{MAHDIZADEH2024543}, making placement decisions repeatable and reproducible by third parties under identical inputs and configurations. The comparative work of centralized and distributed orchestration approaches has demonstrated that the underlying architecture significantly influences workflow completion times and performance variability~\cite{10.1145/3592533.3592809}. A recurring observation across these works is the necessity of empirical characterization of behaviors, such as cold and warm start patterns, queuing latency, and resource footprint—supported by homogeneous benchmarking. Therefore, without such standardized measurements, transferring placement strategies across heterogeneous environments becomes unreliable. These insights motivate our focus on comparable measurements of the same workflow executed in the browser, at the edge, and in the cloud.
\subsection{Cold-start and cost management}
One of the key factors in serverless computing is the \textit{cold start}, referring to the additional latency that occurs when a function invoked without an already warm instance available~\cite{farahani2024serverless}. This forces the platform to provision a new sandbox environment (e.g., microVM or container), load and initialize the runtime along with its dependencies, and bring the function code to a ready-to-execute state. For Wasm, this process includes module validation, compilation, and instantiation. 
However, recent work~\cite{golec2024cold} shows that although cold start can sometimes dominate end-to-end latency in certain production workloads, many functions are not latency-sensitive. Consequently, the authors argue that serverless systems should be evaluated and designed with a broader perspective, considering not only startup latency but also predictability, throughput, and resource efficiency. With respect to cost, recent work~\cite{10.1007/978-3-031-53227-6_32} investigates how pricing models, billing granularity, and data egress fees affect the total cost of serverless applications. These works introduce parameterized evaluation methodologies that facilitate normalized comparisons across providers, yet require workload and runtime measurements for proper calibration.

\subsection{Contributions beyond the state-of-the-art}
Prior works evaluated Wasm runtimes, serverless orchestration, and cold-start behaviors largely in isolation, focusing on a single environment (browser, edge, or cloud) or on micro-benchmarks. To our knowledge, no unified, cross-environment work has executed the same Wasm-based workflow across these three layers under a common experimental setup and set of metrics. This paper instead provides an end-to-end comparison of Wasm-based serverless workflows executed in the browser, at the edge, and in the cloud, offering new insights into how runtime configurations (AOT, JIT) and environmental factors jointly affect performance and resource efficiency across the computing continuum.

\section{Experimental Design}
\label{sec:methodology}
This section explains how we evaluate the impact of the execution environment (browser, edge, and cloud instances), on the performance of Wasm-based serverless workflows under a consistent measurement setup.

\emph{Workflow artifacts}:
We model the application as a directed acyclic graph (DAG), where each node represents a Wasm module with explicit input/output interfaces. The same artifacts are executed across all environments, allowing us to attribute observed performance differences solely to environmental factors rather than to code or packaging variations.

\begin{figure}[t]
\centering
    \includegraphics[width=0.46\textwidth]{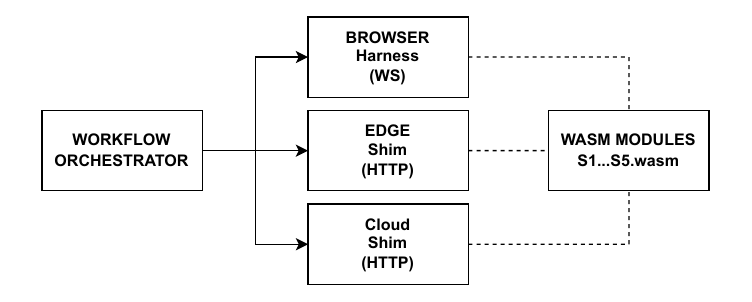}   
    \caption{High-level execution pipeline of the serverless workflow across browser, edge, and cloud environments.}  
    \label{fig:architecture}
\end{figure}

\emph{Workflow orchestrator}:
We developed an external component that serves as a workflow orchestrator, interpreting the DAG and triggering each task according to its structure (e.g., sequence, fan-out, and fan-in). Each task’s output is passed as input to its successors, ensuring that tasks remain isolated and do not invoke one another directly. This design preserves identical flow control across all three environments. The orchestrator also records timing and resource utilization metrics, enabling consistent and comparable measurements across experiments.
The orchestrator runs in a control environment isolated from the workflow Wasm executors to prevent interference. In the browser setup, it is co-located on the same machine but runs in a separate process, communicating via a loopback WebSocket with a harness page that forwards messages to the Web Worker. The Web Worker executes the workflow tasks in a separate thread.
At the edge, the orchestrator runs on a different machine within the same LAN as the executor node and invokes tasks through a lightweight HTTP shim co-located with the Wasm runtime. In the cloud configuration, it runs on a virtual machine (VM) in the same region, using the same HTTP shim interface to trigger executions. Fig~\ref{fig:architecture} illustrates the resulting high-level software architecture.
%%%
%%%

\textit{Metrics and data collection}
We organize the collected metrics into two complementary perspectives for evaluating application performance. The first captures end-to-end workflow behavior and results, while the second focuses on resource utilization and runtime-level effects within the platform:

\textit{(i) User-level metrics}:
    \begin{itemize}
        \item \textit{Cold start}: invocation without an existing warm instance, requiring sandbox provisioning (e.g., container), artifact loading, runtime initialization, and dependency setup;
        \item \textit{Warm start}: served by an already initialized instance without reloading or reinitialization;
        \item \textit{Function Latency}: Start/end timestamps per step;
        \item \textit{Workflow makespan}: total elapsed time from acquiring the first step’s input to producing the final output, including internal queuing and inter-step data transfers.
    \end{itemize}

\textit{(ii) System-level metrics}:
    \begin{itemize}
        \item \textit{Throughput}: number of completed workflow invocations per unit time under steady-state conditions with fixed concurrency;
        \item \textit{Overhead and resources}: CPU (average and peak utilization), memory (RSS and peak), I/O activity, and artifact size (Wasm binary and AOT cache), with the sampling period and measurement tools specified.
    \end{itemize}

\textit{Analysis and reproducibility:} We adopt a robust statistical analysis, reporting median, 25th/75th percentiles (p25/p75), interquartile range (IQR), and, where applicable, \qty{95}{\percent} confidence intervals for key performance indicators. To ensure fair comparisons, we explicitly specify the varying factors between measurements (e.g., execution environment or ablation parameters) while keeping all other parameters constant. In addition, reproducibility is guaranteed through strict artifact versioning, fixed runtime configurations, and deterministic random seeds.

\section{Evaluation Setup}
\label{sec:setup}
This section explains how we executed experiments across browsers, edge, and cloud instances under identical conditions, i.e., Wasm artifact, invocation Application Binary Interface (ABI), and workflow orchestrator. To accommodate browser constraints, the workflow is designed to operate without file-system access. The ABI remains identical across all environments; only the transport mechanism differs: WebSocket with \texttt{postMessage} in the browser, and HTTP \texttt{multipart/form-data} at the edge and in the cloud instances.

Each workflow task is implemented in Rust and compiled into Wasm modules using the same toolchain (Rust $\rightarrow$ wasm32-wasi), allowing AOT and JIT compilation to be toggled as ablation parameters. All builds and corresponding module images are version-controlled via commit hashes or container digests for reproducibility. Compilation flags and enabled Wasm features are kept identical across the environments for consistent execution behavior.
%%%%%%%%%%%%%%%%%%
\subsection{Workflow}
\begin{figure}[t]
\centering
    \includegraphics[width=0.4\textwidth]{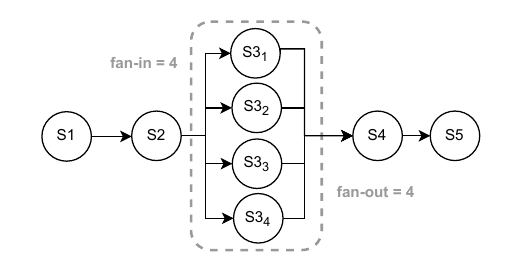}   
    \caption{Workflow DAG.}   
    \label{fig:workflow-example}
    \Description[Workflow DAG]{Workflow DAG consisting of five steps.}
\end{figure}

We evaluate a serverless workflow, modeled as a DAG with ($N=5$) functions and fan-out/fan-in \(\langle f_{\mathrm{out}}, f_{\mathrm{in}}\rangle=\langle 4,4\rangle\), as shown in Fig. ~\ref{fig:workflow-example}.
As mentioned, each step is implemented in Rust and compiled to \texttt{wasm32-wasi} modules that exchange data frames encoded in the Concise Binary Object Representation (CBOR) format entirely in memory.
The workflow processes deterministic synthetic payloads (generated with a fixed seed) in three input sizes:
\(s_{\mathrm{small}}=16\,\mathrm{KiB}\),
\(s_{\mathrm{medium}}=1\,\mathrm{MiB}\),
\(s_{\mathrm{large}}=4\,\mathrm{MiB}\).
The workflow functions are:

\begin{enumerate}[(S1)]
    \item \emph{Ingest\&Deserialize (serialization-bound):} decodes the CBOR frame in memory buffer, validates the schema, and calculates a CRC32 of the input for a quick integrity check. Complexity cost: $\mathcal{O}(n)$.  

    \item \emph{Preprocess (memory-bound):} applies normalization and a scan over the buffer (sliding window, clamp, type conversion \texttt{u8}\(\to\)\texttt{f32}); the output is a buffer of the same size. Complexity cost: $\mathcal{O}(n)$;

    \item \emph{Map (CPU-bound, fan-out=4):} divides the buffer into four blocks and processes them in parallel with a blocked \emph{dense matrix multiplication}. Complexity cost: $\mathcal{O}(d^3)$, with \(d=\sqrt{s/4}\);

    \item \emph{Reduce (serialization-bound, fan-in=4):} aggregates the four results (sum/concatenation according to the scheme) into a single block and calculates a partial digest (\emph{BLAKE3}) for verification. Complexity cost: $\mathcal{O}(n)$;

    \item \emph{Serialize\&Finalize (serialization-bound):} recodes the final output in CBOR and produces the final digest (\emph{BLAKE3}) to be compared with the expected value. Complexity cost: $\mathcal{O}(n)$.
 
\end{enumerate}

\paragraph{CPU kernel sizing (S3)}
The size of the matrix multiplication is derived from the payload:
\[
d(s) = 
\begin{cases}
64 & \text{if } s=s_{\mathrm{small}} \ (\approx 64^2 \cdot 4\,\mathrm{B} \simeq 16\,\mathrm{KiB}),\\
512 & \text{if } s=s_{\mathrm{medium}} \ (\approx 512^2 \cdot 4\,\mathrm{B} \simeq 1\,\mathrm{MiB}),\\
1024 & \text{if } s=s_{\mathrm{large}} \ (\approx 1024^2 \cdot 4\,\mathrm{B} \simeq 4\,\mathrm{MiB}).\\
\end{cases}
\]
Each S3\([k]\) processes a separate block in parallel (fan\text{-}out \(=4\)); S4 combines the results.

% \paragraph{I/O contract and correctness}
% The payload for each invocation is a CBOR map with the fields: \texttt{buf} (byte array) and metadata (\texttt{size}, \texttt{chunking} optional).
% The correctness of each run is accepted only if the invariants and digests match; otherwise, the run is discarded. Specifically, it is verified with: (i) CRC32 in S1; (ii) range limits after S2; (iii) comparison of the final \emph{BLAKE3} digest in S5 with an expected value derived from the seed and parameters. A single violation lead to the run being discarded.

\begin{table*}[t]
\centering
\footnotesize
\caption{Platforms and runtimes.}
\label{tab:platforms}
\setlength{\tabcolsep}{4pt}
\begin{tabularx}{\linewidth}{lllllll}
\toprule
\textbf{Env.} &
\textbf{Host / Instance} &
\textbf{CPU} &
\textbf{RAM} &
\textbf{OS} &
\textbf{Browser/Container} &
\textbf{Wasm Runtime} \\
\midrule
Browser &
\texttt{Workstation} \texttt{(linux/amd64)} &
\texttt{6 cores} &
\texttt{16\,GiB} &
\texttt{Ubuntu\,24.04} &\texttt{Firefox\,143.0.4} &
\texttt{Firefox WebAssembly engine (SpiderMonkey)} \\
 & & & & \\
Edge &
\texttt{Raspberry\,Pi\,4} \texttt{(linux/arm64)} &
\texttt{4 cores} &
\texttt{4\,GiB} &
\texttt{Ubuntu\,24.04} & Docker\,\texttt{28.5.1} (cgroups v2) &
\texttt{WasmEdge\,v0.13.5} \\
 & & & & \\
Cloud &
\texttt{VM} \texttt{(linux/amd64)} &
\texttt{4\,vCPU} &
\texttt{24\,GiB} &
\texttt{Ubuntu\,24.04} & Docker \texttt{28.5.1} &
\texttt{WasmEdge\,v0.13.5} \\
\midrule
\multicolumn{6}{l}{\textit{Wasm features}: WASI \texttt{preview1} (edge/cloud; \texttt{JS shim}, subset in browser); SIMD \texttt{on}; Threads \texttt{on}.} \\
\multicolumn{6}{l}{\textit{Resource limits}: Edge/Cloud via cgroups (\texttt{-\--cpus}, \texttt{-\--memory}); Browser via Web~Worker isolation.} \\
\bottomrule
\end{tabularx}
\end{table*}

\subsection{Platforms, Runtimes, and Tooling}

To ensure consistency across browsers, edges, and clouds, we use the same Wasm artifact (target wasm32-wasi, identical toolchain and flags) and fix runtime versions and configurations. The invocation details are described in Section~\ref{sec:execution-env}; here we summarize the platforms, runtimes, and tools used. Table~\ref{tab:platforms} lists, for each environment, hardware/OS, runtime/browser versions, Wasm flags, and resource limits.

\paragraph{Browser host}
The machine running the browser is an Ubuntu 24.04 LTS workstation with an Intel® Core i7-8700K (6 cores, 12 threads, 3.70 GHz) and 16 GiB RAM. We use Mozilla Firefox 143.0.4 (64-bit, Snap for Ubuntu), with WebAssembly SIMD enabled. The workflow is executed in a Web Worker, while the orchestrator runs in a separate process on the same machine and communicates via loopback WebSocket with a harness page.

\paragraph{Edge Node}
The edge runs the same module inside a minimal OCI/Docker container with WasmEdge v0.13.5. The host is Ubuntu 24.04 LTS (linux/arm64) with Docker Engine 28.5.1 (cgroups v2); resources are capped via container limits (-\--cpus 4, -\--memory 4). Wasm features (SIMD) are aligned with the browser; warm state is achieved through a pool of pre-initialized instances, while cold state is forced by resetting the pool. No disk I/O in the workflow; messages pass through memory to the executor.

\paragraph{Cloud VM}
We replicate the edge setup on a linux/amd64 VM co-located with the orchestrator (same region): 4 vCPU, 24 GiB RAM, Ubuntu 24.04; container runtime Docker Engine 28.5.1 (containerd v1.7.28, runc 1.3.0). The Wasm executor is WasmEdge v0.13.5, with the same toolchain configuration as on the edge. Resource limits and cold/warm policies are identical to edge to make the results comparable. Again, the workflow does not access the file system; orchestration uses in-memory messages.

\subsection{Execution Environments}\label{sec:execution-env}

We adopt a unified invocation ABI for all environments. Each task receives a request frame using CBOR serialization for metadata fields. On edge/cloud, large binary payloads are transmitted separately via HTTP multipart/form-data to avoid a base64 or similar encoding overhead within the CBOR envelope. This hybrid approach maintains protocol type safety while supporting payloads up to several megabytes without buffer overflow issues. The minimum structure of exchanged messages is as follows:

\begin{itemize}
    \item request: \texttt{\{op:"invoke", step\_id, run\_id, payload\}}
    \item response: \texttt{\{op:"result", status, payload[, error]\}}
\end{itemize}

The orchestrator sends/waits for these frames and correlates the measurements via \texttt{run\_id}. Since we do not use the file system in the workflow, input/output travels as buffers in memory.

\paragraph{Browser}
The orchestrator communicates with a harness page via WebSocket. The page forwards the frames to the Web Worker with transferable \texttt{ArrayBuffer} \texttt{postMessage} and sends the Web Worker's response back to the orchestrator. Cold start: new Web Worker instance and cache/Service Worker invalidation; warm start: Web Worker and module already compiled remain alive. Timestamps are collected with \texttt{performance.now()}.

\paragraph{Edge} 
The orchestrator invokes a minimal HTTP endpoint on the edge node: \texttt{POST /invoke} (CBOR binary body). A shim in the container delivers the frame to the Wasm executor (WasmEdge) via pipe/IPC and returns the response frame. Cold start: one-shot mode; warm start: pool of pre-initialized instances.

\paragraph{Cloud} 
Same edge pipeline on VM/container in the same region as the orchestrator: POST /invoke (CBOR binary body, HTTP/1.1) and same cold/warm behavior as above. Timestamps are recorded with a monotonic clock (Rust \texttt{Instant}) and the shim returns a CBOR response including the phase breakdown (load, compile, instantiate, init, execute), total latency, and best-effort CPU/RSS samples collected from \texttt{/proc}.

\subsection{Measurement Protocol}

In the experiments, we use a mixed set of micro and macro workloads. The micro workloads include:
(i) CPU-bound (e.g., matmul) with synthetic payloads of size
\(\langle s_{\mathrm{small}}, s_{\mathrm{medium}}, s_{\mathrm{large}} \rangle\);
(ii) memory-bound (allocation/scan);
(iii) serialization-bound (CBOR \(\leftrightarrow\) in\--memory).
The macro workflow is a DAG with \(\langle N \rangle\) steps, fan-out/fan-in
\(\langle f_{\mathrm{out}}, f_{\mathrm{in}} \rangle\), and explicit I/O scheme; for each step, we specify the format and expected size of input/output.

The measurements of latency, cold/warm behavior, and throughput are performed in a closed-loop configuration.
Each configuration (environment \(\times\) payload \(\times\) mode ablation) is repeated \(k\) times \((k=\text{20}\) with randomized order, separate warm-up, and fixed seed.
Timestamps are collected with monotonic clocks (browser: \texttt{performance.now()}; edge/cloud: high-resolution clock) and correlated via trace-id/span-id;
we record: cold-start breakdown \((\mathrm{load}\!\rightarrow\!\mathrm{compile/AOT\text{-}load}\!\rightarrow\!\mathrm{instantiate}\!\rightarrow\!\mathrm{init}\!\rightarrow\!\mathrm{run})\), per-step and end-to-end latencies, workflow makespan, throughput, CPU (average/peak), and memory (RSS/peak).

Outlier, retry, and exclusions follow predefined rules: we discard a run if the monitor detects abnormal conditions;
for valid samples, we apply an outlier filter based on \(1.5\times\mathrm{IQR}\).
\section{Experimental Results}
\label{sec:result}
In this section, we present the experimental results obtained by running the same WebAssembly workflow separately in the browser, edge, and cloud. The analyses are organized along two complementary perspectives: \textit{User-Perceived Performance} and \textit{System-Level Metrics}. The results are aggregated over \(k\) repetitions.

Before discussing performance, we quantify the size of the artifacts. Table \ref{tab:artifact_sizes_mb_percentage} compares, for each workflow module, the size of the Wasm bytecode and that of the AOT (precompiled) output, also reporting the AOT/Wasm percentage increase; this provides the context for interpreting the benefits observed in cold tests with AOT and for estimating the impact of deployment in the three environments.

\begin{table}[t]
\centering
\caption{Artifact sizes comparison between WASM modules and their AOT compiled outputs.}
\label{tab:artifact_sizes_mb_percentage}
\small
\setlength{\tabcolsep}{8pt}
\begin{tabularx}{\linewidth}{lccc}
\toprule
\textbf{Step} &
\makecell{\textbf{Size WASM}\\\textbf{Module (MB)}} &
\makecell{\textbf{Size AOT}\\\textbf{Artifact (MB)}} &
\makecell{\textbf{Size}\\\textbf{Increase}} \\
\midrule
S1 & 0.152 & 0.342 & + 125\% \\
S2 & 0.135 & 0.317 & + 135\% \\
S3 & 0.141 & 0.322 & + 128\% \\
S4 & 0.151 & 0.354 & + 134\% \\
S5 & 0.153 & 0.360 & + 135\% \\
\bottomrule
\end{tabularx}
\end{table}

\subsection{User-Perceived Performance}
We analyze the application's performance that directly impacts the user experience, using three predefined metrics: (i) cold/warm startup times, (ii) latencies per workflow step, and (iii) end-to-end makespan.

\begin{figure}[t]
\centering
    \includegraphics[width=0.45\textwidth]{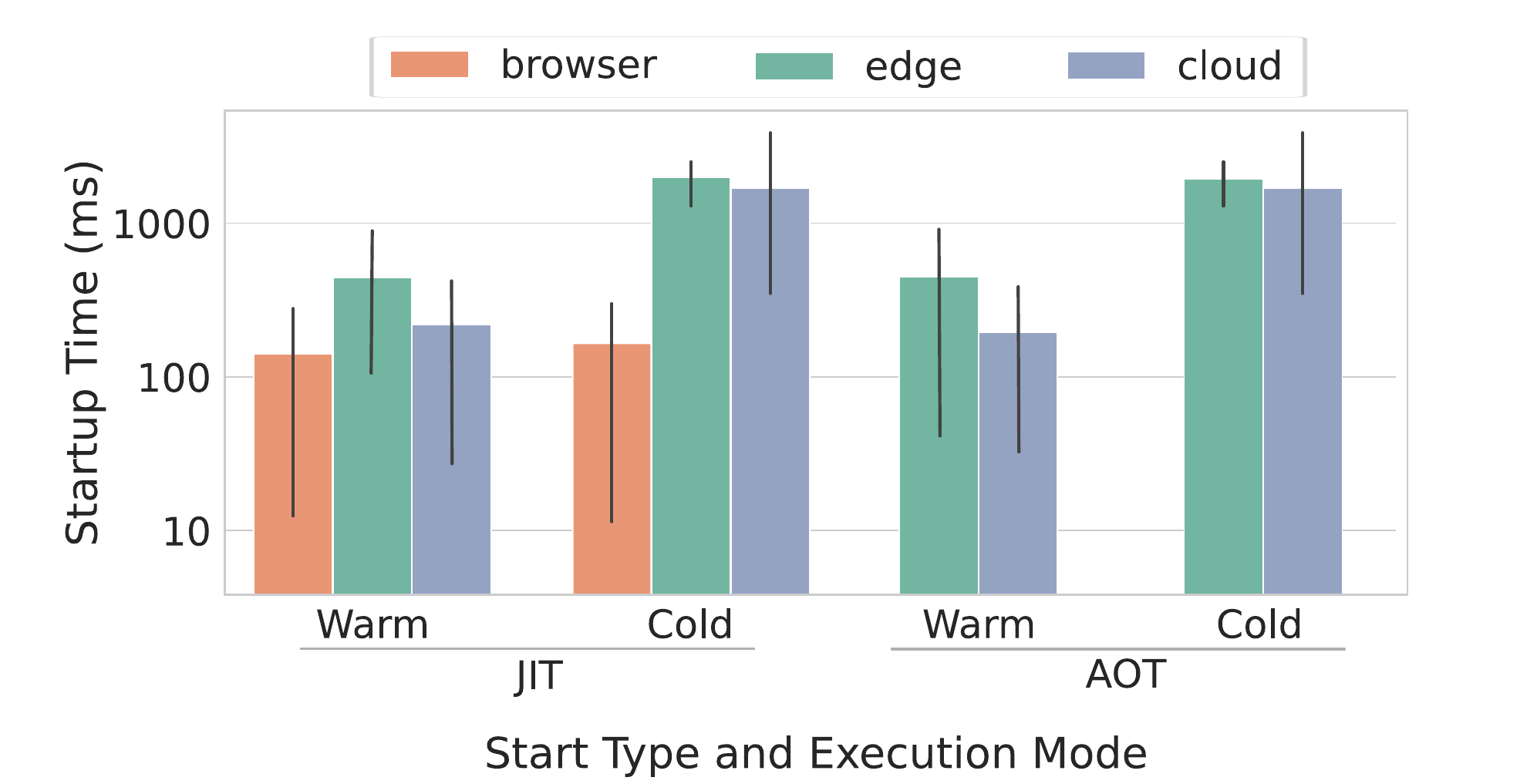}   
    \caption{Cold vs Warm startup times.}   
    \label{fig:startup-times}
\end{figure}

\paragraph{Cold and warm startup}
As shown in Figure \ref{fig:startup-times}, warm starts are faster than cold starts in all environments, although this gap is less pronounced in browsers. The use of AOT, which is only applicable on edge/cloud, significantly reduces the amount of time spent on compilation/validation in cold tests. The browser shows particularly low warm start times, benefiting from compilation/instance already residing in the Worker; edge shows the highest cost in cold JIT, while cloud ranks in the middle with slightly lower variability.

\begin{figure}[t]
\centering
    \includegraphics[width=0.38\textwidth]{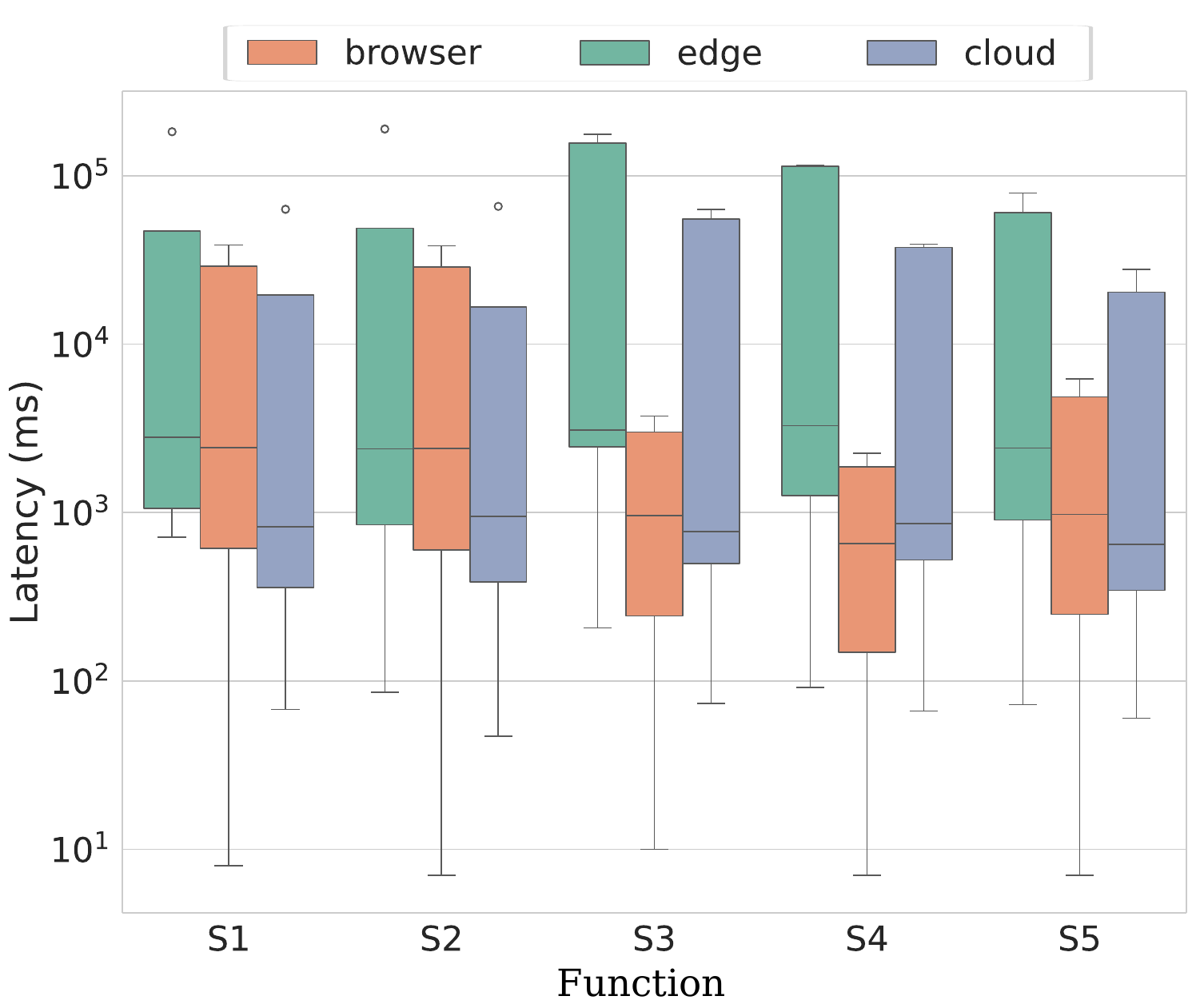}   
    \caption{Function latency.}   
    \label{fig:step-latencies}
\end{figure}

\paragraph{Function latency}
We run the workflow with a warm profile and 1MiB as payload size and collect the start/end timestamps for each step (S1–S5). In Figure \ref{fig:step-latencies} we represent the distributions with boxplots on a logarithmic scale: box = IQR (p25–p75), center line = median; whisker = 1.5×IQR; outliers shown explicitly. The browser performs better on S3–S4–S5 (lower medians and narrower IQRs), thanks to in-memory execution and lower orchestration/IPC overhead. On S1–S2, it ranks between edge and cloud: the overhead of (de)serialization and the Worker-main thread transition (postMessage/buffer transfer) make it less advantageous for the browser, while edge/cloud runtimes with optimized containers and I/O pipes mitigate the cost of these steps. In summary, when computation or aggregation prevails (S3–S4–S5), the browser is competitive; when (de)serialization dominates (S1–S2), the advantage diminishes.

\begin{figure}[t]
\centering
    \includegraphics[width=0.38\textwidth]{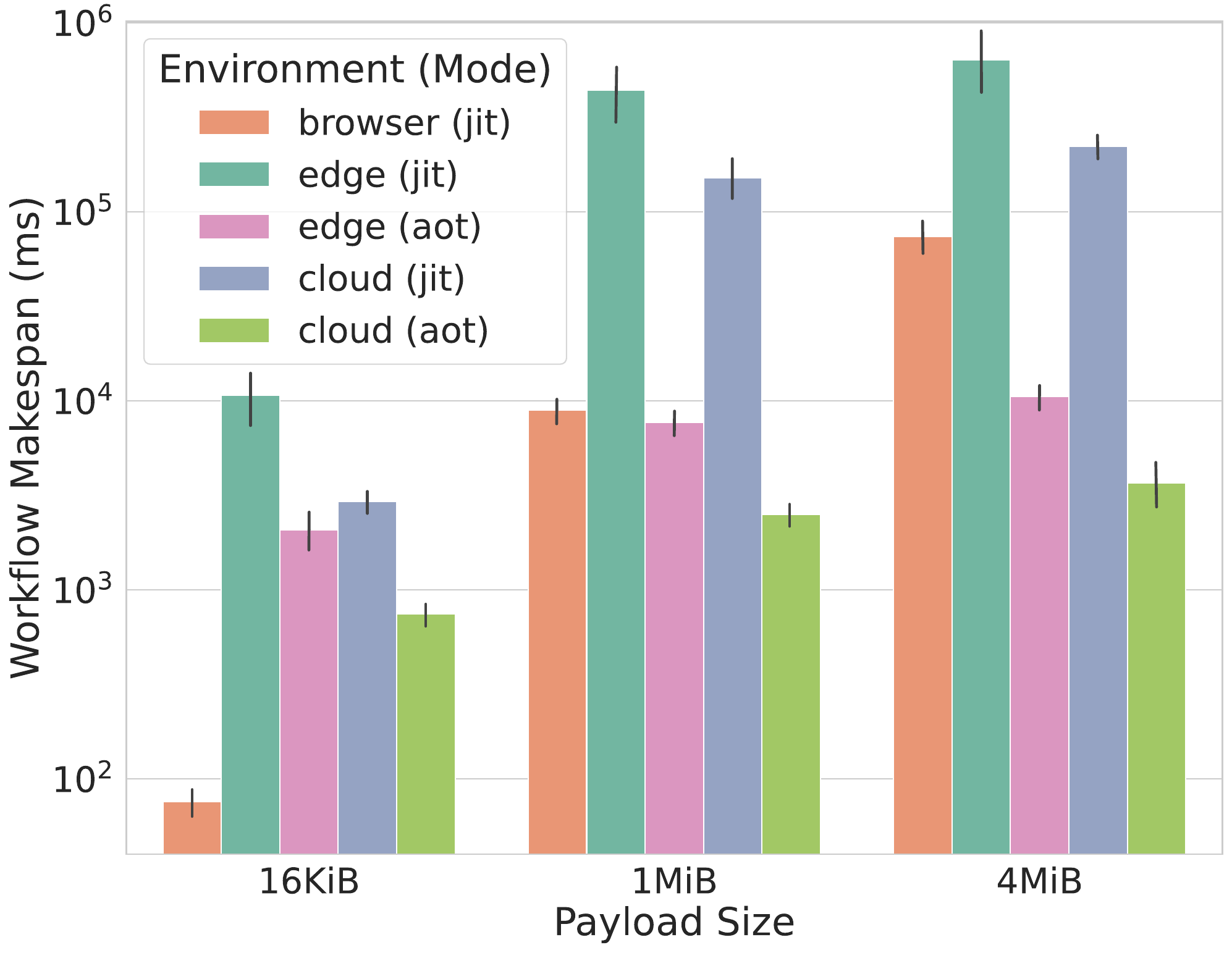}   
    \caption{Workflow makespan.}   
    \label{fig:makespan}
\end{figure}

\paragraph{Workflow makespan}
We measure the time S1→S5 (including queues and transfers) in \textit{warm} mode by varying the \textit{payload size} and the \textit{JIT/AOT} mode. In Figure \ref{fig:makespan}, we observe a clear reversal as the payload increases: with small sizes, the browser (JIT) is often faster because the startup cost is minimal, everything happens in-memory, and Worker orchestration has little impact. Moving to medium/large sizes, the makespan becomes compute/memory-bound: the startup cost is amortized and memory bandwidth, cache, and code quality prevail; here, AOT configurations on edge/cloud are faster and more stable. In particular, edge AOT tends to emerge when the load is markedly numerical, while the browser's advantage diminishes until it reverses. In all cases, warm and AOT improve predictability and end-to-end times, with more pronounced effects as size increases.

\subsection{System-Level Metrics}
This section observes the platform's behavior under load, complementing user-facing metrics. We consider sustainable capacity (throughput) and resource usage profile (CPU/RSS), measured with the same experimental setup.

\begin{figure}[t]
\centering
    \includegraphics[width=0.38\textwidth]{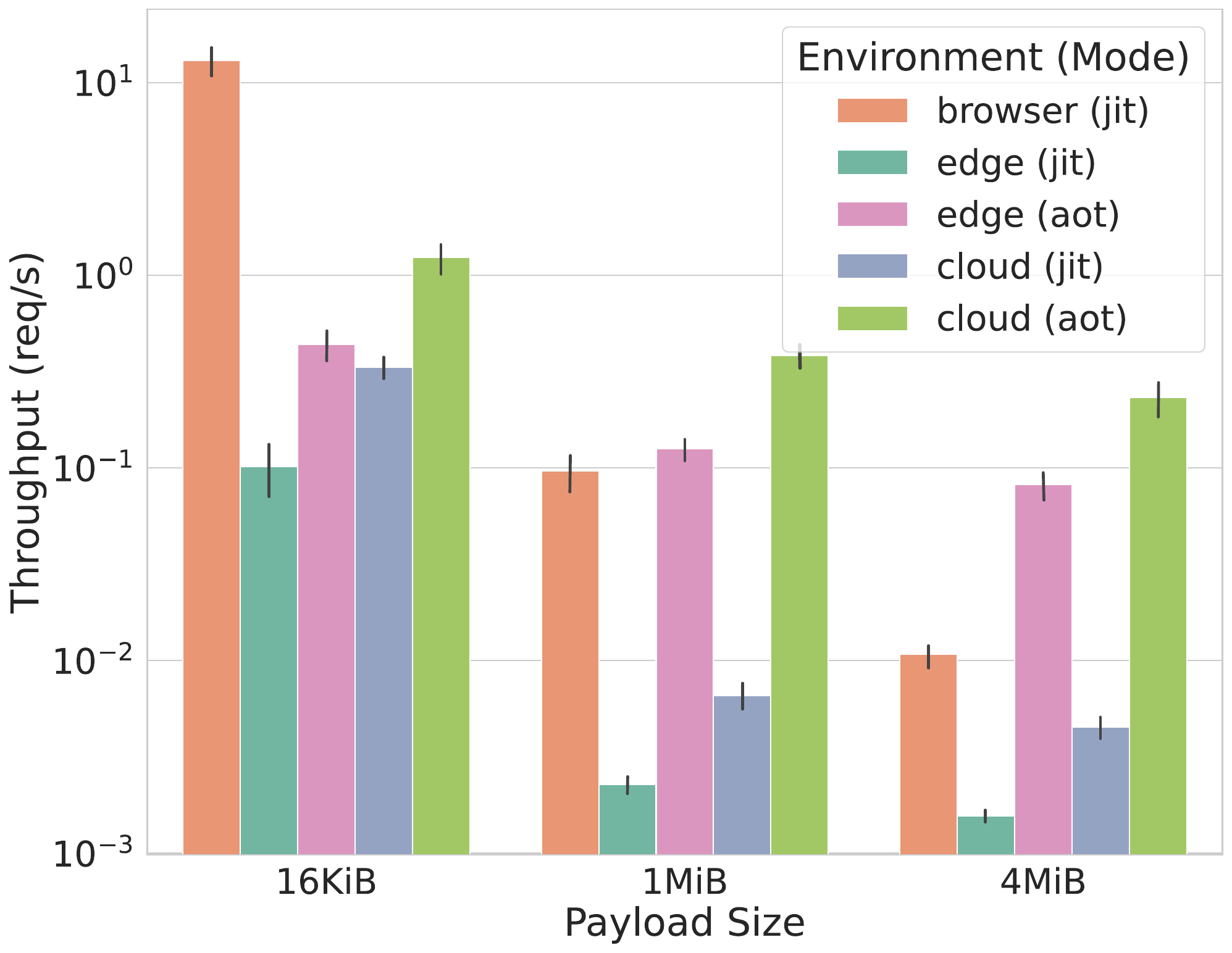}   
    \caption{Workflow throughput.}   
    \label{fig:throughput}
\end{figure}

\paragraph{Throughput} A new invocation is issued upon completion of the previous one, so that the reported value corresponds to the inverse of the average completion time. Figure \ref{fig:throughput} shows results consistent with those for makespan. With 16 KiB, the browser (JIT) prevails: lightweight invocations, in-memory exchanges, and reduced startup costs allow for tens of requests per second. At 1 MiB and 4 MiB, the compute/memory-bound effect emerges: the initial overhead is amortized, and AOT configurations on edge/cloud outperform the browser due to the elimination of JIT compilation and more favorable CPU/memory-bandwidth/cache.

\begin{figure}[t]
\centering
    \includegraphics[width=0.45\textwidth]{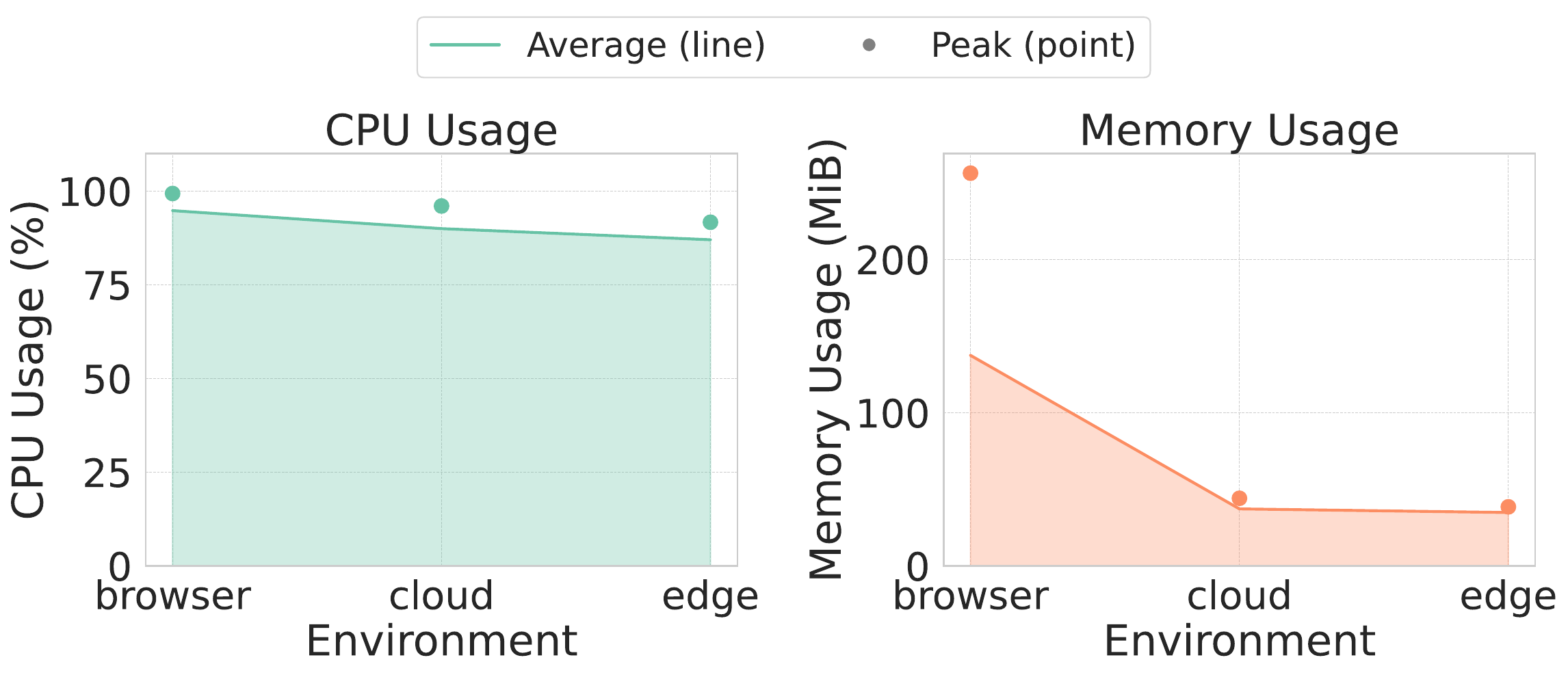}   
    \caption{Resource usage.}   
    \label{fig:resources}
\end{figure}

\paragraph{Resource usage}
We sample the CPU (average/peak) and memory (average/peak RSS) of the executor process, with a sampling interval of 20 ms on edge/cloud (readings from /proc) and 20 ms in the browser. In the browser, CPU is obtained by recording a Performance session from in-browser DevTools and aggregating the profile samples; the same tools on the Web Worker side estimate memory. As shown in Figure \ref{fig:resources}, CPU usage is high in all environments (indicating computational bottlenecks), with average values slightly lower on the edge than in the cloud and browser. Memory usage differs more significantly: the browser has higher average/peak RSS (JS engine stack, Worker, GC management), while the cloud and edge remain more compact thanks to the native Wasm runtime and the absence of browser components.

\section{Conclusion}
\label{sec:conclusion}
We presented a comparative analysis of the execution of the same serverless workflow in browsers, edge, and cloud, keeping Wasm and ABI artifacts unchanged. The results indicate that warm and AOT significantly reduce startup times and variability. The browser is competitive on small loads and on steps dominated by orchestration/aggregation, while edge/cloud in AOT prevail with increasing payloads when the regime becomes compute/memory-bound. Makespan and throughput consistently reflect these trends, and CPU/RSS profiles explain the differences. This evidence should be interpreted within the scope of the setup (runtimes, hardware, browser, workload) and does not claim to be generalizable beyond the experiment.

% As future work, we intend to refine the initialization and execution phases and expand runtime and browser coverage. We will explore hybrid workflows that span the entire browser–edge–cloud continuum, by dynamically placing steps or subgraphs based on payload size, fan-out, and operating conditions, including offloading mechanisms (from browser to edge/cloud and vice versa) to balance latency, capacity, and resource utilization.
\section*{Acknowledgment}
This work received partial funding from the Italian Ministry of University and Research (MUR) ``Research projects of National Interest (PRIN-PNRR)'' call through the project ``Cloud Continuum aimed at On-Demand Services in Smart Sustainable Environments'' (CUP: J53D23015080001- IC: P2022YNBHP), and "SEcurity and RIghts in the CyberSpace (SERICS)" project (PE00000014), under the MUR National Recovery and Resilience Plan funded by the European Union - NextGenerationEU (CUP: D43C22003050001), the European Union’s Horizon Europe research and innovation program, grant agreements 101093202 (Graph-Massivizer), and the Austrian Research Promotion Agency (FFG, grant agreement 909989, AIM AT Stiftungsprofessur für Edge AI)

\bibliographystyle{ACM-Reference-Format}
\bibliography{main}
\end{document}